\documentclass[aps,prl,superscriptaddress,amsmath,amssymb,noshowkeys,a4paper,twocolumn,showpacs]{revtex4}
\usepackage[utf8]{inputenc}

\usepackage{amsmath}
\usepackage{amssymb,amsfonts,latexsym}
\usepackage{graphicx}
\usepackage{epsfig}
\usepackage{color}
\usepackage{textcomp}
\usepackage{comment}
\usepackage{epstopdf}

\begin{document}
\title{Discontinuous Buckling of Wide Beams and Metabeams}
\author{Corentin Coulais}
\affiliation{Huygens-Kamerlingh Onnes Lab, Universiteit Leiden, PObox 9504, 2300 RA Leiden, The Netherlands}
\affiliation{FOM Institute AMOLF, Science Park 104, 1098 XG Amsterdam, The Netherlands}
\author{Johannes T. B. Overvelde}
\affiliation{School of Engineering and Applied Sciences, Harvard university, Cambridge, Massachusetts 02138, USA}
\author{Luuk A. Lubbers}
\affiliation{Huygens-Kamerlingh Onnes Lab, Universiteit Leiden, PObox 9504, 2300 RA Leiden, The Netherlands}
\affiliation{FOM Institute AMOLF, Science Park 104, 1098 XG Amsterdam, The Netherlands}
\author{Katia Bertoldi}
\affiliation{School of Engineering and Applied Sciences, Harvard university, Cambridge, Massachusetts 02138, USA}
\author{Martin van Hecke}
\affiliation{Huygens-Kamerlingh Onnes Lab, Universiteit Leiden, PObox 9504, 2300 RA Leiden, The Netherlands}
\affiliation{FOM Institute AMOLF, Science Park 104, 1098 XG Amsterdam, The Netherlands}

%
%


\begin{abstract}
We uncover how nonlinearities dramatically alter the buckling of
elastic beams. First, we show experimentally that sufficiently wide ordinary elastic beams
and specifically designed metabeams ---beams made from a mechanical metamaterial--- exhibit discontinuous buckling, an unstable form of buckling
where the post-buckling stiffness is negative. Then we use simulations to uncover the crucial role of nonlinearities,
and show that beams made from increasingly nonlinear materials exhibit increasingly negative post-buckling slope.
Finally,  we
demonstrate that for sufficiently strong nonlinearity, we can observe
discontinuous buckling for metabeams as slender as $1\%$ numerically and $5\%$
experimentally.
\end{abstract}
\maketitle

%

\begin{figure}[b!]
\includegraphics[width=\columnwidth]{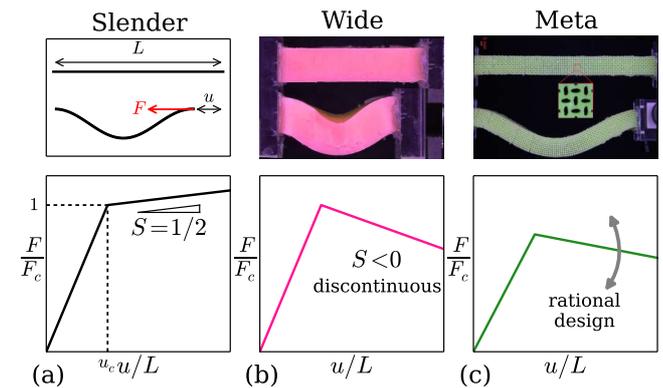}
\caption{(color online). Buckling of slender, wide and metabeams. (a) Slender beams in their
undeformed (top) and buckled (bottom) states. The force displacement curve for slender beams has a post-buckling slope
$S=1/2$.  (b) Discontinuous $(S<0)$ buckling of
wide beams.
(c) Slender metabeams consisting of a nonlinear elastic metamaterial can also
exhibit discontinuous buckling.}
\label{fig:meta_sketch}
\end{figure}

 Buckling of straight  beams under sufficient  load $F$ is perhaps the
most basic example of an elastic instability.
This instability can be captured in models of varying degree of
sophistication~\cite{Truskinovsky_2007}, starting from
Euler's elastica which describes the bending
of elastic lines and is appropriate for slender beams~\cite{Euler_1744}. Even the simplest analysis gives an excellent estimate of the critical buckling load $F_c$, crucial for
engineering~\cite{bazantbook}. The relation between $F$ and compressive displacement $u$ for a beam of length $L$ takes the form
$(F-F_c)/F_c = S ~u/L$, with the elastica predicting that
the post-buckling slope $S$ equals  $1/2$, independent of boundary conditions ---see Fig.~\ref{fig:meta_sketch}a.

Here we describe how nonlinearities ---due to large strains in wide beams or due to
strong nonlinearities in  metabeams--- dramatically alter this post-buckling scenario.
In particular we find that sufficiently strong nonlinearities lead to
{\em discontinuous buckling}, a novel form of buckling
where the force in the post-buckling regime {\em decreases} for increasing deformation, so that $S<0$.

First, we perform experiments on ordinary elastic beams and show that they undergo discontinuous buckling
when the beams aspect ratio exceeds $12\%$ ---see Fig.~\ref{fig:meta_sketch}b. Second, we create beams out of a
strongly nonlinear mechanical metamaterial, and show that such beams undergo discontinuous buckling
when the materials nonlinearity becomes sufficiently strong, even for slender beams ---see Fig.~\ref{fig:meta_sketch}c.
Third, using finite element simulations, we uncover
a significant nonlinear contribution to the total elastic energy of both wide beams and metabeams,
which we suggest pushes the beam away from the Euler limit and
causes discontinuous buckling. To test this hypothesis, we numerically study a range of metabeams and show that
the strength of the nonlinearity of their stress-strain relation and
their postbuckling slopes are strongly correlated.
Finally,
we present numerical evidence that a judicious choice of metamaterial parameters can cause arbitrarily slender beams to
exhibit discontinuous buckling, and experimentally achieve discontinuous buckling for metabeams as slender as $5\%$.
Our work illuminates the crucial role of nonlinearities for buckling, and paves the way for novel strategies where
mechanical metamaterials are used to qualitatively change and control the nature of elastic instabilities.

{\em Discontinuous Buckling:} We first perform experiments on the buckling of both ordinary elastic beams and metabeams.  To minimize
gravitational effects we perform density-matched experiments in a bath of water. We rigidly mount the beams (using silicon glue) to the top and bottom plate
of an Instron 5965 uniaxial testing device equipped with a $100$~N load cell, allowing us to measure the axial force $F$ (accuracy $10^{-4}$~N)
as a function of the axial compressive displacement $u$ (accuracy $10^{-3}$~mm).

The wide beams are solid, rectangular beams of length $L=45$~mm, depth $d=35$~mm and widths ranging from $w=1.55$ mm
to $w=12.85$~mm (aspect ratios, or thicknesses, $t:=w/L$ up to $0.27$)~\cite{NoteExp1}. These are created by molding
a well-characterized silicon rubber~\cite{elite 8}. The metabeams
consist of a rubber mechanical metamaterial designed to allow
tuning of the effective stress-strain nonlinearity.
We take inspiration from a recently proposed  mechanical metamaterial, which consists of a 2D elastic slab
patterned by a regular array of circular holes. Such system exhibits an elastic instability under compression
leading to a transformation to a pattern of mutually orthogonal ellipses and a sharp kink in the stress-strain
relation~\cite{Mullin_PRL2007,Bertoldi_AdvM2010,Bertoldi_JMPS2008,
Overvelde_AdvM2012}.
Here we use instead metamaterials with elliptical holes (Fig.~2c),
which break rotational symmetry and suppress this elastic instability~\cite{Taylor_AdvMat2014,Florijn_PRL2014},
transforming the sharp kink into a controllable nonlinearity of the stress-strain relation.
We have created  six  metabeams ($L=220$ mm, $d=29$ mm, $w=24$ mm, $t=0.10$ and $E=1.1\times 10^6$~Pa)
with a varying aspect ratio between the ellipses by 3d printing molds
in which we cast a silicon rubber~\cite{metapar}. Each beam contains $9\times98=882$
holes, where the strongest nonlinearities occur for near-circular holes.

\begin{figure}[t!]
\includegraphics[width=1.\columnwidth]{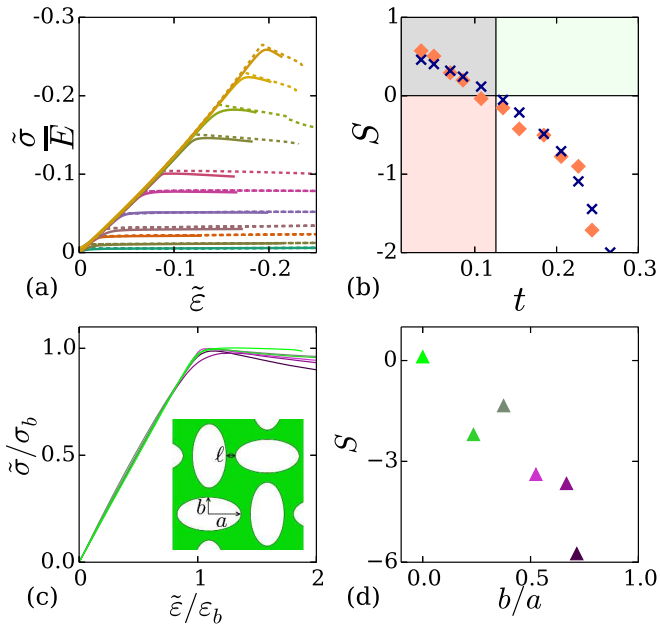}
\caption{(color online). Discontinuous buckling. (a) Force-displacement curves for
beams with aspect ratios ranging from $t=0.034$ (lowest curve)
to $t=0.284$ (highest curve) for experiments (solid) and finite elements simulations (dashed).
Here,  $\tilde{\sigma}/E:= -F/(E w d)$ and $\tilde{\varepsilon}:= -u/L$.
(b) The post-buckling slope in experiments (orange diamonds) and simulations
(blue crosses), $S$, increasingly deviates from the elastica limit $1/2$ for large $t$. Discontinuous buckling ($S<0$)
occurs for $t>0.12$.  (c)
Force-displacement curves (rescaled by the buckling stress and strain, $\sigma_b$ and $\varepsilon_b$) for metabeams of  $t=0.10$, $\ell=0.3$ and various
   values of $b/a$. Here
 $\sigma_b$ and $\varepsilon_b$ denote the values at buckling.  Inset: geometry of our metamaterial.
   (d) Corresponding post-buckling slope $S$ vs. $b/a$.}
  \label{fig:beamssetup}
  \end{figure}
We plot the experimental force-displacement curves and the
post-buckling slope $S$ for wide beams in Fig.~\ref{fig:beamssetup}ab, and for metabeams in Fig.~\ref{fig:beamssetup}cd.
For all beams, there is an initially near-linear elastic behavior with a sudden departure from linearity as a result of buckling~\cite{NoteSulcus}.
Our first main result is that for {ordinary beams with} $t\gtrsim0.12$, or for metabeams with near-circular holes, the post-buckling
slope becomes {\em negative}.
Hence, under increasing load, such beams exhibit {\em discontinuous
buckling}. Note that a negative stiffness is readily observed in other
mechanical systems such as buckling shells~\cite{bazantbook}, the Roorda frame and pipes~\cite{hutchinson1970postbuckling},
where it is associated with asymmetric or saddle node bifurcations. Together 
with wrinkling membranes~\cite{Pocivavsek_Science2008,Diamant_PRL2011,Audoly_PRE2011}, this 
example is one of the few where a negative stiffness 
is reported for a mechanical system undergoing a symmetry breaking pitchfork bifurcation.


{\em Numerical Approach:} In order to understand discontinuous buckling,
we make extensive use of finite element simulations and conduct a fully non-linear analysis within 
the commercial package Abaqus/Standard. To calibrate the constitutive relation,
we first focus on wide, ordinary beams. These undergo substantial uniaxial compression before buckling, pushing the physics beyond that of simple linear elasticity.
Such rubber-like materials are well described
by the incompressible neo-Hookean formulation of elasticity, which leads to a nonlinear stress-strain relation~\cite{Boyce2000,ogden}.
In Fig.~\ref{fig:beamssetup}a we compare our experimental data to finite element simulations
of such a neo-Hookean $3D$ model, with realistic (fixed) boundary conditions,
for $E=250$ kPa and $\nu=0.49999$~\cite{NoteProtocol}, and find excellent agreement between experiments and
simulations, validating the use of this weakly nonlinear model.
In the remainder of the paper we will use 2D (plain strain) simulations~\cite{NotePLaneStrain}: (i) of the full metabeam to extract $S$;
(ii) of a unit cell with periodic boundary conditions to determine the effective stress-strain relation for a uniaxial test.



{{\em Nonlinearity:} We now illustrate and quantify the role of nonlinearity for the stresses and elastic energies in the post-buckling regime, comparing
three beams:  a slender ordinary beam close to the Euler limit, a wide beam and a metabeam.
In Figs.~\ref{fig:profiles}a-c we show the effective stress-strain relation of these beams
(extracted from our numerical simulations), as well as the range of axial strains and stresses throughout the whole
3D slender and wide beams, and throughout the whole 2D metabeam.
To facilitate comparison of the strength of the nonlinearities, all data is taken at  $\tilde{\varepsilon}/\varepsilon_b=120\%$,
where $\varepsilon_b$ denotes the onset of buckling.
Figs.~\ref{fig:profiles}a and \ref{fig:profiles}b illustrate that while for a slender beam  ($t=0.034$) the strains only span a limited range ($\Delta\tilde{\varepsilon}=1.2\times10^{-3}$) so 
that the stresses are not very sensitive to the neo-Hookean nonlinearity, for a wide beam ($t=0.134$) the strains span a larger range ($\Delta\tilde{\varepsilon}=1.8\times10^{-1}$) and the stresses 
thus deviate significantly from the linear (Euler) case. Moreover,
Fig.~\ref{fig:profiles}c illustrates that metabeams with a strongly nonlinear stress-strain relation exhibit stresses that deviate 
significantly from the linear case even for small strains.
Clearly both the width (setting the range of strains) and the nonlinearity of the material 
(setting the curvature of the stress-strain relation) play a role in determining the deviations from the Euler limit.

To quantify the role of nonlinearity, we will now determine the contributions to the elastic energy of bending, compression,
and nonlinearity in the regime close to the buckling strain $\varepsilon_b$. To do so, we need to determine the constitutive law as well as an equation for
the axial strain as function of $x$, the horizontal coordinate across the beam width $w$.
For the constitutive law we expand the stress-strain relation to quadratic order around $\varepsilon_b$:
\begin{equation}
\tilde{\sigma}_{zz}=E(\tilde{\varepsilon}_{zz}+\eta(\tilde{\varepsilon}_{zz}-\varepsilon_b)^2)~,\label{eqeta}
\end{equation}
where $\eta$  quantifies the nonlinearity \cite{footnotezero}.
The axial strain profile is expanded as\begin{equation}
\tilde{\varepsilon}_{zz} = \varepsilon + \kappa x,\label{eq:profile}
\end{equation}
where $\kappa$ and $\varepsilon$ are respectively the curvature and the compression of the neutral plane of the beam.
Neglecting shear (which can be shown to be subdominant \cite{luukCCinpreparation}),  the elastic energy can then be determined as
$E_t=\int\textrm{d}V \int \tilde{\sigma}_{zz} \textrm{d}\tilde{\varepsilon}_{zz}=
E_c+E_b+E_{NL}$, with
\begin{eqnarray}
E_b &=&\!\frac{E d w^3}{12}\!\int_0^L\!\textrm{d}s\,\kappa^2~, \label{eq:bending} \\
E_c &=&\!E d w\!\int_0^L\!\textrm{d}s\,\varepsilon^2~,\label{eq:compression}\\
E_{NL}&=&\!\eta\frac{E d w}{3} \!\int_0^L\!\textrm{d}s\!\left(\!\frac{w^2}{4}\!(\!\varepsilon\!-\!\varepsilon_b\!)
\kappa^2\!+\!(\!\varepsilon\!-\!\varepsilon_b\!)^3\!+\!\varepsilon_b^3\!\right)\! \label{eq:NL}~,
\end{eqnarray}
where $s$ is the curvilinear coordinate of the beam. We note that the elastica only uses $E_b$, whereas extensible
elastica uses both $E_b$ and $E_c$, but does not take nonlinearities, such as those encountered in
neo-Hookean materials into account~\cite{Boyce2000,ogden} --- consequently, 
the post-buckling slope in such models remains positive up to unrealistically large 
aspect ratios~\cite{Magnusson_IJSS2011,humer2013}. We have recently developed
a full theoretical description taking $E_{NL}$ into account, which
is quantitatively consistent with our experimental
and numerical data, and which will appear separately~\cite{luukCCinpreparation}.

\begin{figure}[t!]
    \includegraphics[width=\columnwidth]{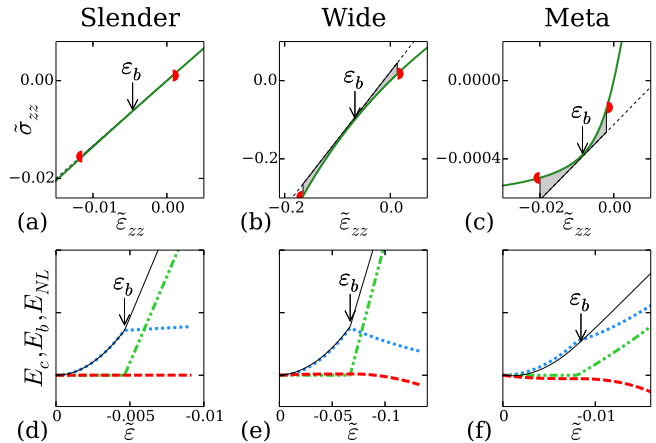}
\caption{
(color online). (a-c) Stresses in 3D slender (a: $L\times d = 45\textrm{mm} \times 35\textrm{mm}$, $t=0.034$) and wide
(b: same $L\times d$, $t=0.134$) beams and for a 2D metabeam (c: $t=0.054$).
The green curves depict the stress-strain relation, the black dashed lines denote linear fits
around $\tilde{\varepsilon_b}$, and the red semicircles denote the range of stresses and strains
throughout the beams for $\tilde{\varepsilon}/\tilde{\varepsilon_b}=1.2$.
(d-f) Energies $E_c$ (dotted, blue), $E_b$ (dot-dashed, green) and $E_{NL}$ (dashed, red)
(Eqs.~(\ref{eq:bending}-\ref{eq:NL})) and total energy $E_t$ (black) vs.
displacement $\tilde{\varepsilon}$. A quadratic fit to $E_t$ in the post-buckling regime gives $\partial^2 E_t/\partial \tilde{\varepsilon}^2=1.6\times10^{-2}$ (a),
$-1.1\times10^{-1}$ (b) and $-4.4\times10^{-2}$ (c).}
\label{fig:profiles}
\end{figure}

Here we focus on comparing the contributions of these energies for slender,
wide and metabeams, and extract $\kappa$ and $\varepsilon$ from the neutral plane/line
of the simulated beams (Figs.~\ref{fig:profiles}d-f).
For the slender beam, the nonlinear term remains small, and
after buckling, $E_t$ grows faster than linear so
that $S:=-\left(E F_c\right)^{-1}\partial \tilde{\sigma}/\partial \tilde{\varepsilon}= \left(1/F_c\right)\partial^2 E_t/\partial \tilde{\varepsilon}^2 $ is positive (Fig.~\ref{fig:profiles}d), as expected.
In contrast, for
the wide beam, the nonlinear contribution becomes significant and induces a sublinear increase of the total
energy, leading to $S<0$ (Fig.~\ref{fig:profiles}e). For metabeams, the nonlinear contribution becomes
similarly important and leads to $S<0$ also.

We suggest that the significant nonlinear contribution upsets the energy balance and
perturbs the beam away from the slender beam limit. Figs.~\ref{fig:profiles}d-f illustrate
the opposite nature of the changes in compressive energy between wide beams ($\eta<0$) and metabeams ($\eta >0$).
Wide beams lower their energy by extending after buckling (in contrast to slender beams), 
due to the neo-Hookean nonlinearity which is stiffening under compression; 
Metabeams lower their energy by shortening more than slender beams after buckling, due to the
constitutive nonlinearity which is softening under compression.
In both cases, stronger nonlinearities lead to an increasing deviation from the Euler limit, leading to a change in the beam geometry and eventually
to discontinuous buckling.



{\em Tunable Nonlinearity in Metabeams: } To establish the connection between the nonlinearity of the metamaterial and the post-buckling slope of the metabeams, we perform extensive simulations of our 2D homogeneous metamaterials and metabeams, scanning the meta-parameters  $\ell$ and $e:= 1-b/a$ as well as beam thickness $t$.
Fig.~\ref{fig:meta_beam_quest} compares $\eta$ and $S$ for a range of $\ell$ and $e:= 1-b/a$, for beams of $t=0.054$. Clearly a smaller gap between the holes $\ell$ leads to larger nonlinearities, whereas the trend with $e$ is non-monotonic. Crucially, the data
shows a strong correspondence between $\eta$ and $S$, which confirms that for given $t$,
the strength of the nonlinearity is the essential parameter which sets the post-buckling slope, and that a judicious choice of the meta-parameters can lead to strongly discontinuous buckling.

\begin{figure}[t!]
\includegraphics[width=\columnwidth]{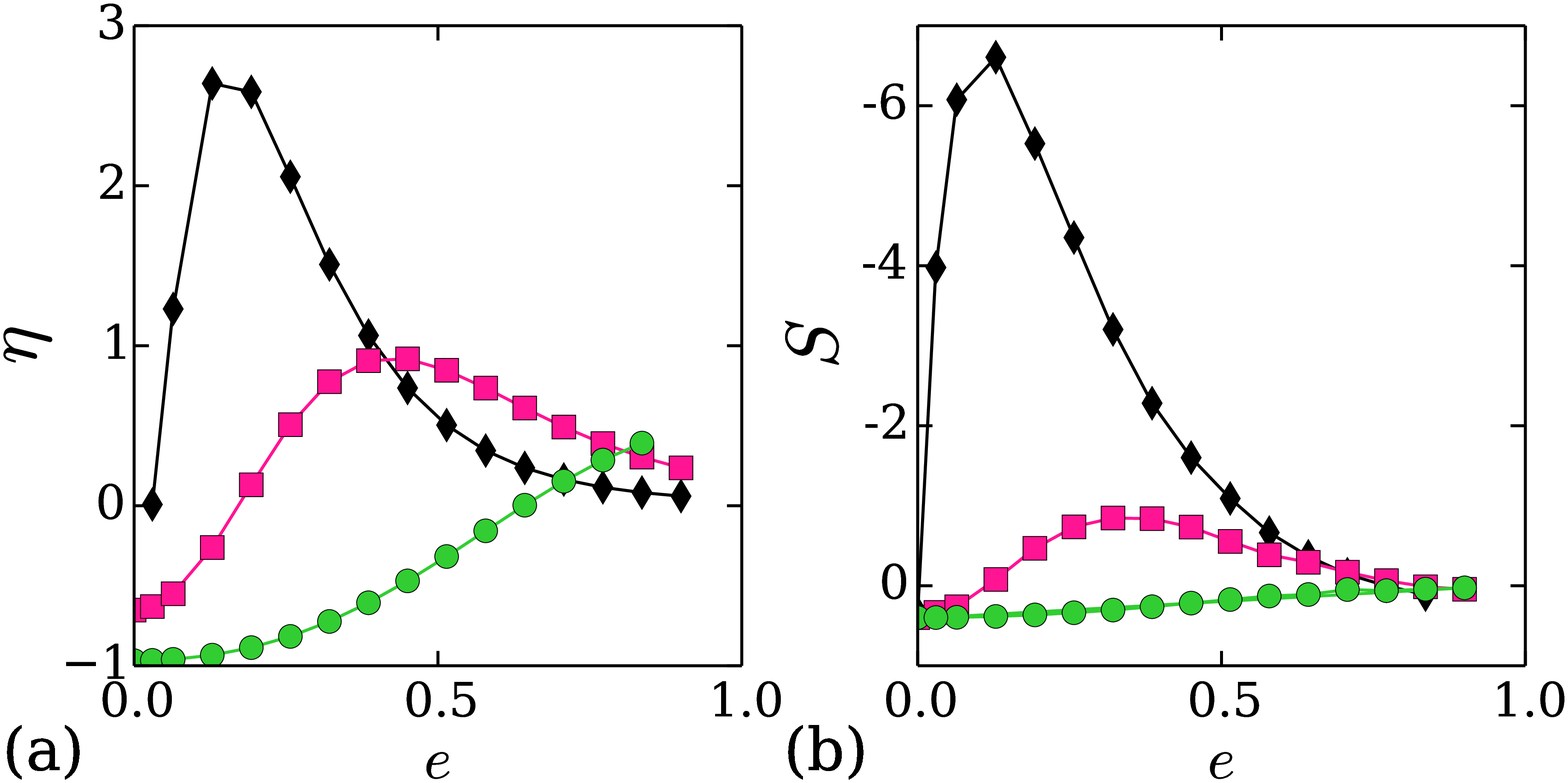}
\caption{ (color online). Numerical simulations of 2D metabeams of $5.4\%$ aspect ratio ($\varepsilon_b=-0.0096$)
    and $6\times 111$ holes for the nonlinearity $\eta$ (a) and post-buckling slope $S$ (b) vs. $e$
for $\ell=0.13$ (black diamonds), $\ell=0.23$ (pink squares) and $\ell=0.42$ (green circles). See~\cite{suppmatMovie} for movies.}
\label{fig:meta_beam_quest}
\end{figure}

{\em Discontinuous Buckling of Slender Beams:} Our scenario suggests that slender beams can exhibit discontinuous buckling when $\eta$ becomes sufficiently large.
We have numerically determined the boundary in the $\ell-e$ plane
between positive and negative $S$ for beams of thicknesses ranging from $9\%$ to
$1\%$ (Fig.~\ref{fig:MB_rational}a).
As expected, to exhibit discontinuous buckling, thinner beams require smaller values of $\ell$, tantamount to stronger nonlinearities.
Pushing our computational power to the edge, we find numerical examples of
$t=0.01$ beams that exhibit discontinuous buckling --- here $\eta \approx
70$! Crucially, our data indicates that the critical value of $\ell$ scales
linearly with $t$, so that suitable chosen metabeams
can exhibit discontinuous buckling for arbitrary small values of the
slenderness.

We also used this data to rationally design an experimental metabeam with
desired post-buckling behavior.
We pick a specific set of metaparameters ($e=0.1$,
$\ell=0.2$) for which our numerics indicates that discontinuous buckling occurs for a critical aspect ratio
$t \gtrsim$ 5\% ---see Fig.~\ref{fig:MB_rational}a. We 3D print a mold consisting of $6
\times 330$ pillars (pitch = $1.65$~mm) with these parameters, and mold a beam
of length $520$ mm, width $9.5$~mm and $1980$ holes ---see Figs.~\ref{fig:MB_rational}b-d.
By lateral clamping we vary the effective length $L_e$ 
of the metabeam, and thus its effective aspect ratio $t_e:=9.5\textrm{mm} / L_e$~\cite{NoteBCS}.
Fig.~\ref{fig:MB_rational}e shows that  discontinuous buckling
sets in for $t_e \gtrsim 5$\%, illustrating the success of our design
strategy.


\begin{figure}[t!]
\includegraphics[width=1.\columnwidth]{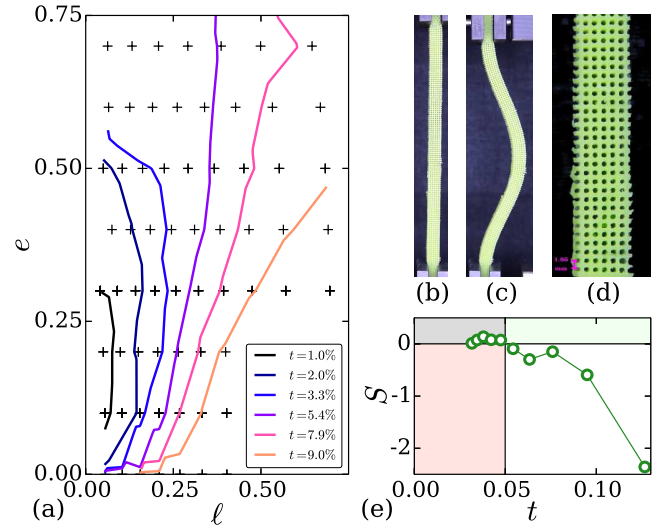}
\caption{(color online). Rational design of discontinuous buckling.
(a) Boundaries between ordinary (right) and discontinuous (left) buckling in the ($e$,$\ell$,$t$) parameter space. For each pair of ($e$,$\ell$) values (crosses), simulations for a range of  beam thicknesses have been performed.
(b-d) Large metabeam ($L=520$ mm, $w=9.5$ mm, $d=16.0$ mm, $E=1.1\times 10^6$ Pa)
with $1980$ holes with $e \approx 0.1$ and $\ell \approx 0.2$.
(e) Experimental post-buckling slope $S$ vs. aspect ratio $t$.}
  \label{fig:MB_rational}
\end{figure}

{\em Discussion and Outlook:} In this work, we showed how nonlinearity can alter the post-buckling mechanics of buckling: when 
the product of $|\eta|$ and critical strain is large enough, nonlinearities lead to discontinuous buckling. Whereas
Euler theory is asymptotically valid for sufficiently linear materials when the thickness tends to zero, none of its 
current extensions~\cite{Magnusson_IJSS2011,humer2013} actually predicts such qualitative change of the post-instability for realistic aspect ratios.

Our strategy is generic and opens up pathways for the rational design of other mechanical phenomena. We expect 
that this approach could be used to design the 2D buckling patterns~\cite{audoly2010elasticity} of metaplates. Could the snapping 
instability used in micro-actuators and sensors ~\cite{Brenner19082003} be tweaked?
Could we design metamaterials for which the post-buckling stiffness is larger than $1/2$? Finally, we note that most mechanical metamaterials have a
beam-like microstructure~\cite{Wegener_reviewRPP2008,Kadic_APL2012,Oftadeh_PRL2014,Florijn_PRL2014}, and often draw on buckling for their
functionality~\cite{Mullin_PRL2007,Bertoldi_JMPS2008,Bertoldi_AdvM2010,Shim_PNAS2012,Babaee_AdvM2013,Overvelde_JMPS2014,Wang_PRL2014}.
We envision that tunable microscopic buckling will be of great use for the rational design of hierarchical metamaterials~\cite{Oftadeh_PRL2014,Cho_PNAS2014,Lakes_Nature1993}.

{\em Acknowledgements}. We thank J. Mesman for outstanding technical support. We acknowledge E.-J. Vegter, J. Lugthart, R. Bastiaansen for
exploratory experiments and theory, K. Kamrin, S. Neukirch and S. Waitukaitis for discussions, NWO/VICI for funding, SEAS for hospitality.


\begin{thebibliography}{21}
\expandafter\ifx\csname natexlab\endcsname\relax\def\natexlab#1{#1}\fi
\expandafter\ifx\csname bibnamefont\endcsname\relax
  \def\bibnamefont#1{#1}\fi
\expandafter\ifx\csname bibfnamefont\endcsname\relax
  \def\bibfnamefont#1{#1}\fi
\expandafter\ifx\csname citenamefont\endcsname\relax
  \def\citenamefont#1{#1}\fi
\expandafter\ifx\csname url\endcsname\relax
  \def\url#1{\texttt{#1}}\fi
\expandafter\ifx\csname urlprefix\endcsname\relax\def\urlprefix{URL }\fi
\providecommand{\bibinfo}[2]{#2}
\providecommand{\eprint}[2][]{\url{#2}}


\bibitem[{\citenamefont{Truskinovsky}(2007)}]{Truskinovsky_2007}
\bibinfo{author}{\bibfnamefont{Y.}~\bibnamefont{Grabovsky}} \bibnamefont{and}
  \bibinfo{author}{\bibfnamefont{L.}~\bibnamefont{Truskinovsky}},
  \emph{\bibinfo{title}{The flip side of buckling}},
  \bibinfo{journal}{Continuum Mechanics and Thermodynamics},
  \textbf{\bibinfo{volume}{19}}, \bibinfo{pages}{211--243}
   (\bibinfo{year}{2007}).

 \bibitem[{\citenamefont{Euler}(1744)}]{Euler_1744}
\bibinfo{author}{\bibfnamefont{L.}~\bibnamefont{Euler}},
\emph{\bibinfo{title}{Additamentum I de curvis elasticis, methodus inveniendi lineas curvas maximi minimivi proprietate gaudentes}},
\bibinfo{journal}{Opera Omnia}, \bibinfo{number}{1}, \textbf{\bibinfo{volume}{24}}, \bibinfo{pages}{245-310} (\bibinfo{year}{1744}).



\bibitem[{\citenamefont{Bazant and Cendolin}(2009)}]{bazantbook}
\bibinfo{author}{\bibfnamefont{Z.}~\bibnamefont{Bazant}} \bibnamefont{and}
  \bibinfo{author}{\bibfnamefont{L.}~\bibnamefont{Cendolin}},
  \emph{\bibinfo{title}{Stability of Structures}} (\bibinfo{publisher}{World
  Scientific}, \bibinfo{year}{2009}).


 \bibitem{NoteExp1} We use large $d$ to avoid out of plane buckling.

 \bibitem{elite 8} Polyvinyl Siloxane double elite 8, Young's modulus $E=250$~kPa, Poisson's ratio $\nu\approx 0.5$.



\bibitem[{\citenamefont{Mullin et~al.}(2007)\citenamefont{Mullin, Deschanel,
  Bertoldi, and Boyce}}]{Mullin_PRL2007}
\bibinfo{author}{\bibfnamefont{T.}~\bibnamefont{Mullin}},
  \bibinfo{author}{\bibfnamefont{S.}~\bibnamefont{Deschanel}},
  \bibinfo{author}{\bibfnamefont{K.}~\bibnamefont{Bertoldi}}, \bibnamefont{and}
  \bibinfo{author}{\bibfnamefont{M.~C.} \bibnamefont{Boyce}},
  \emph{\bibinfo{title}{Pattern Transformation Triggered by Deformation}},
  \bibinfo{journal}{Phys. Rev. Lett.} \textbf{\bibinfo{volume}{99}},
  \bibinfo{pages}{084301} (\bibinfo{year}{2007})
  .

\bibitem[{\citenamefont{Bertoldi et~al.}(2010)\citenamefont{Bertoldi, Reis,
  Willshaw, and Mullin}}]{Bertoldi_AdvM2010}
\bibinfo{author}{\bibfnamefont{K.}~\bibnamefont{Bertoldi}},
  \bibinfo{author}{\bibfnamefont{P.~M.} \bibnamefont{Reis}},
  \bibinfo{author}{\bibfnamefont{S.}~\bibnamefont{Willshaw}}, \bibnamefont{and}
  \bibinfo{author}{\bibfnamefont{T.}~\bibnamefont{Mullin}},
  \emph{\bibinfo{title}{Negative Poisson's Ratio Behavior Induced by an Elastic Instability}},
  \bibinfo{journal}{Adv. Mater.} \textbf{\bibinfo{volume}{22}},
  \bibinfo{pages}{361} (\bibinfo{year}{2010}).



\bibitem[{\citenamefont{Bertoldi et~al.}(2008)\citenamefont{Bertoldi, Boyce,
  Deschanel, Prange, and Mullin}}]{Bertoldi_JMPS2008}
\bibinfo{author}{\bibfnamefont{K.}~\bibnamefont{Bertoldi}},
  \bibinfo{author}{\bibfnamefont{M.}~\bibnamefont{Boyce}},
  \bibinfo{author}{\bibfnamefont{S.}~\bibnamefont{Deschanel}},
  \bibinfo{author}{\bibfnamefont{S.}~\bibnamefont{Prange}}, \bibnamefont{and}
  \bibinfo{author}{\bibfnamefont{T.}~\bibnamefont{Mullin}},
  \emph{\bibinfo{title}{Mechanics of deformation-triggered pattern transformations
  and superelastic behavior in periodic elastomeric structures}},
  \bibinfo{journal}{J. Mech. Phys. Solids}
  \textbf{\bibinfo{volume}{56}}, \bibinfo{pages}{2642 } (\bibinfo{year}{2008}).

\bibitem[{\citenamefont{Overvelde et~al.}(2012)\citenamefont{Overvelde, Shan,
  and Bertoldi}}]{Overvelde_AdvM2012}
\bibinfo{author}{\bibfnamefont{J.~T.~B.} \bibnamefont{Overvelde}},
  \bibinfo{author}{\bibfnamefont{S.}~\bibnamefont{Shan}}, \bibnamefont{and}
  \bibinfo{author}{\bibfnamefont{K.}~\bibnamefont{Bertoldi}},
  \emph{\bibinfo{title}{Compaction Through Buckling in 2D Periodic, Soft and Porous Structures: Effect of Pore Shape}},
  \bibinfo{journal}{Adv. Mater.} \textbf{\bibinfo{volume}{24}},
  \bibinfo{pages}{2337} (\bibinfo{year}{2012}).


  \bibitem[{\citenamefont{Florijn et~al.}(2014)\citenamefont{Florijn, Coulais,
  and van Hecke}}]{Florijn_PRL2014}
\bibinfo{author}{\bibfnamefont{B.}~\bibnamefont{Florijn}},
  \bibinfo{author}{\bibfnamefont{C.}~\bibnamefont{Coulais}}, \bibnamefont{and}
  \bibinfo{author}{\bibfnamefont{M.}~\bibnamefont{van Hecke}},
  \emph{\bibinfo{title}{Programmable Mechanical Metamaterials}},
  \bibinfo{journal}{Phys. Rev. Lett.} \textbf{\bibinfo{volume}{113}},
  \bibinfo{pages}{175503} (\bibinfo{year}{2014}).

    \bibitem[{\citenamefont{Taylor et~al.}(2014)\citenamefont{Taylor, Francesconi,
  Gerendás, Shanian, Carson, and Bertoldi}}]{Taylor_AdvMat2014}
\bibinfo{author}{\bibfnamefont{M.}~\bibnamefont{Taylor}},
  \bibinfo{author}{\bibfnamefont{L.}~\bibnamefont{Francesconi}},
  \bibinfo{author}{\bibfnamefont{M.}~\bibnamefont{Gerendás}},
  \bibinfo{author}{\bibfnamefont{A.}~\bibnamefont{Shanian}},
  \bibinfo{author}{\bibfnamefont{C.}~\bibnamefont{Carson}}, \bibnamefont{and}
  \bibinfo{author}{\bibfnamefont{K.}~\bibnamefont{Bertoldi}},
          \emph{\bibinfo{title}{Low Porosity Metallic Periodic Structures with Negative Poisson's Ratio}},
  \bibinfo{journal}{Adv. Mater.} \textbf{\bibinfo{volume}{26}},
  \bibinfo{pages}{2365} (\bibinfo{year}{2014}).


  \bibitem{metapar} Polyvinyl Siloxane double elite 32, Young's modulus $E=1.1$~MPa, Poisson's ratio $\nu\approx 0.5$.

 \bibitem{NoteSulcus} The range over which we can measure the post-buckling slope is limited,
as at larger values of $\tilde{\varepsilon}$ the experimental beams
develop a crease or {\em sulcus}~\cite{citeulike:9431849,PhysRevLett.99.076101,PhysRevLett.106.105702,PhysRevLett.109.025701},
which is outside the scope of this work.



\bibitem[{\citenamefont{Gent}(2005)}]{citeulike:9431849}
\bibinfo{author}{\bibfnamefont{A.~N.} \bibnamefont{Gent}},
\emph{\bibinfo{title}{Elastic instabilities in rubber}},
  \bibinfo{journal}{Int. J. NonLin. Mech.}
  \textbf{\bibinfo{volume}{40}}, \bibinfo{pages}{165} (\bibinfo{year}{2005}).

\bibitem[{\citenamefont{Ghatak and Das}(2007)}]{PhysRevLett.99.076101}
\bibinfo{author}{\bibfnamefont{A.}~\bibnamefont{Ghatak}} \bibnamefont{and}
  \bibinfo{author}{\bibfnamefont{A.~L.} \bibnamefont{Das}},
  \emph{\bibinfo{title}{Kink Instability of a Highly Deformable Elastic Cylinder}},
  \bibinfo{journal}{Phys. Rev. Lett.} \textbf{\bibinfo{volume}{99}},
  \bibinfo{pages}{076101} (\bibinfo{year}{2007}).

\bibitem[{\citenamefont{Hohlfeld and Mahadevan}(2011)}]{PhysRevLett.106.105702}
\bibinfo{author}{\bibfnamefont{E.}~\bibnamefont{Hohlfeld}} \bibnamefont{and}
  \bibinfo{author}{\bibfnamefont{L.}~\bibnamefont{Mahadevan}},
      \emph{\bibinfo{title}{Unfolding the Sulcus}},
  \bibinfo{journal}{Phys. Rev. Lett.} \textbf{\bibinfo{volume}{106}},
  \bibinfo{pages}{105702} (\bibinfo{year}{2011}).

\bibitem[{\citenamefont{Hohlfeld and Mahadevan}(2012)}]{PhysRevLett.109.025701}
\bibinfo{author}{\bibfnamefont{E.}~\bibnamefont{Hohlfeld}} \bibnamefont{and}
  \bibinfo{author}{\bibfnamefont{L.}~\bibnamefont{Mahadevan}},
    \emph{\bibinfo{title}{Scale and Nature of Sulcification Patterns}},
  \bibinfo{journal}{Phys. Rev. Lett.} \textbf{\bibinfo{volume}{109}},
  \bibinfo{pages}{025701} (\bibinfo{year}{2012}).


   \bibitem[{\citenamefont{Hutchinson}(1970)}]{hutchinson1970postbuckling}
\bibinfo{author}{\bibfnamefont{J.~W.} \bibnamefont{Hutchinson}} \bibnamefont{and}
  \bibinfo{author}{\bibfnamefont{W.~T.} \bibnamefont{Koiter}},
\emph{\bibinfo{title}{Postbuckling theory}},
  \bibinfo{journal}{Appl. Mech. Rev.}
  \textbf{\bibinfo{volume}{23}}, \bibinfo{pages}{1353} (\bibinfo{year}{1970}).

 
   \bibitem[{\citenamefont{Pocivavsek et~al}(2008)\citenamefont{Pocivavsek, Luka and Dellsy, Robert and Kern, Andrew and
   Johnson, Sebastián and Lin, Binhua and Lee, Ka Yee C. and Cerda, Enrique}}]{Pocivavsek_Science2008}
\bibinfo{author}{\bibfnamefont{L.}~\bibnamefont{Pocivavsek}},
\bibinfo{author}{\bibfnamefont{R.}~\bibnamefont{Dellsy}},
\bibinfo{author}{\bibfnamefont{A.}~\bibnamefont{Kern}},
\bibinfo{author}{\bibfnamefont{S.}~\bibnamefont{Johnson}},
\bibinfo{author}{\bibfnamefont{B.}~\bibnamefont{Lin}},
\bibinfo{author}{\bibfnamefont{K. Y. C.}~\bibnamefont{Lee}}, \bibnamefont{and}
  \bibinfo{author}{\bibfnamefont{E.}~\bibnamefont{Cerda}},
  \emph{\bibinfo{title}{Stress and Fold Localization in Thin Elastic Membranes}},
  \bibinfo{journal}{Science} \textbf{\bibinfo{volume}{320}},
  \bibinfo{pages}{912-916} (\bibinfo{year}{2008}).

   \bibitem[{\citenamefont{Diamant and Witten}(2011)\citenamefont{Diamant and Witten}}]{Diamant_PRL2011}
\bibinfo{author}{\bibfnamefont{H.}~\bibnamefont{Diamant}} \bibnamefont{and}
  \bibinfo{author}{\bibfnamefont{T. A.}~\bibnamefont{Witten}},
  \emph{\bibinfo{title}{Compression Induced Folding of a Sheet: An Integrable System}},
  \bibinfo{journal}{Phys. Rev. Lett.} \textbf{\bibinfo{volume}{107}},
  \bibinfo{pages}{164302} (\bibinfo{year}{2011}).

%
  \bibitem[{\citenamefont{Audoly}(2011)\citenamefont{Audoly}}]{Audoly_PRE2011}
\bibinfo{author}{\bibfnamefont{B.}~\bibnamefont{Audoly}},
  \emph{\bibinfo{title}{Localized buckling of a floating elastica}},
  \bibinfo{journal}{Phys. Rev. E} \textbf{\bibinfo{volume}{84}},
  \bibinfo{pages}{011605} (\bibinfo{year}{2011}).


\bibitem[{\citenamefont{Boyce and Arruda}(2000)}]{Boyce2000}
\bibinfo{author}{\bibfnamefont{M.~C.} \bibnamefont{Boyce}} \bibnamefont{and}
  \bibinfo{author}{\bibfnamefont{E.~M.} \bibnamefont{Arruda}},
  \emph{\bibinfo{title}{Constitutive Models of Rubber Elasticity: A Review}},
  \bibinfo{journal}{Rubber Chemistry and Technology}
  \textbf{\bibinfo{volume}{73}}, \bibinfo{pages}{504} (\bibinfo{year}{2000}).

\bibitem[{\citenamefont{Ogden}(1997)}]{ogden}
\bibinfo{author}{\bibfnamefont{R.~W.} \bibnamefont{Ogden}},
  \emph{\bibinfo{title}{Non Linear Elasic Deformations}}
  (\bibinfo{publisher}{Dover Publ}, \bibinfo{year}{1997}).
  
\bibitem{NoteProtocol} For details about the numerical scheme, see Supplemental Information.

  \bibitem{NotePLaneStrain} The 2D plane strain approximation describes our 3D experimental 
 situations, where the depth of the beam is much larger than its width.


  \bibitem{footnotezero} Clearly this expansion cannot be extended down to $\tilde{\varepsilon}=0$.

  \bibitem[{\citenamefont{Lubbers et~al.}(2014)\citenamefont{Lubbers, Coulais,
  and van Hecke}}]{luukCCinpreparation}
\bibinfo{author}{\bibfnamefont{L. A.}~\bibnamefont{Lubbers}},
  \bibinfo{author}{\bibfnamefont{C.}~\bibnamefont{Coulais}}, \bibnamefont{and}
  \bibinfo{author}{\bibfnamefont{M.}~\bibnamefont{van Hecke}},
      \emph{\bibinfo{title}{Elastica for Nonlinear Beams}},
  \bibinfo{journal}{(in preparation)}  (\bibinfo{year}{2015}).


\bibitem{suppmatMovie} For movies, see supplemental information.


\bibitem{NoteBCS} Such clamping boundary conditions are not exactly the same as the no-slip
boundary conditions used before.


  \bibitem[{\citenamefont{Magnusson et~al.}(2001)\citenamefont{Magnusson, Ristinmaa, and
  Ljung}}]{Magnusson_IJSS2011}
\bibinfo{author}{\bibfnamefont{A.} \bibnamefont{Magnusson}},
  \bibinfo{author}{\bibfnamefont{M.} \bibnamefont{Ristinmaa}}, \bibnamefont{and}
  \bibinfo{author}{\bibfnamefont{C.} \bibnamefont{Ljung}},
            \emph{\bibinfo{title}{Behaviour of the extensible elastica solution}},
  \bibinfo{journal}{Intl. J. Solids Struct.}
  \textbf{\bibinfo{volume}{38}}, \bibinfo{pages}{8441} (\bibinfo{year}{2001}).
  

\bibitem[{\citenamefont{Humer}(2013)}]{humer2013}
\bibinfo{author}{\bibfnamefont{A.}~\bibnamefont{Humer}},
    \emph{\bibinfo{title}{Exact solutions for the buckling and postbuckling of shear-deformable beams}},
\bibinfo{journal}{Acta   Mech.} \textbf{\bibinfo{volume}{224}}, \bibinfo{pages}{1493}
  (\bibinfo{year}{2013}).


\bibitem[{\citenamefont{Audoly and Pomeau}(2010)}]{audoly2010elasticity}
\bibinfo{author}{\bibfnamefont{B.}~\bibnamefont{Audoly}} \bibnamefont{and}
  \bibinfo{author}{\bibfnamefont{Y.}~\bibnamefont{Pomeau}},
  \emph{\bibinfo{title}{Elasticity and geometry}} (\bibinfo{publisher}{Oxford
  Univ. Press}, \bibinfo{year}{2010}).

\bibitem[{\citenamefont{Brenner et~al.}(2003)\citenamefont{Brenner, Lang, Li,
  Qiu, and Slocum}}]{Brenner19082003}
\bibinfo{author}{\bibfnamefont{M.~P.} \bibnamefont{Brenner}},
  \bibinfo{author}{\bibfnamefont{J.~H.} \bibnamefont{Lang}},
  \bibinfo{author}{\bibfnamefont{J.}~\bibnamefont{Li}},
  \bibinfo{author}{\bibfnamefont{J.}~\bibnamefont{Qiu}}, \bibnamefont{and}
  \bibinfo{author}{\bibfnamefont{A.~H.} \bibnamefont{Slocum}},
            \emph{\bibinfo{title}{Optimal design of a bistable switch}},
  \bibinfo{journal}{Proc. Natl. Ac. Sci. U.S.A.}
  \textbf{\bibinfo{volume}{100}}, \bibinfo{pages}{9663} (\bibinfo{year}{2003}).



  \bibitem[{\citenamefont{Kadic et~al.}(2012)\citenamefont{Kadic, B\"{u}ckmann,
  Stenger, Thiel, and Wegener}}]{Kadic_APL2012}
\bibinfo{author}{\bibfnamefont{M.}~\bibnamefont{Kadic}},
  \bibinfo{author}{\bibfnamefont{T.}~\bibnamefont{B\"{u}ckmann}},
  \bibinfo{author}{\bibfnamefont{N.}~\bibnamefont{Stenger}},
  \bibinfo{author}{\bibfnamefont{M.}~\bibnamefont{Thiel}}, \bibnamefont{and}
  \bibinfo{author}{\bibfnamefont{M.}~\bibnamefont{Wegener}},
  \emph{\bibinfo{title}{On the practicability of pentamode mechanical metamaterials}},
  \bibinfo{journal}{Appl. Phys. Lett.} \textbf{\bibinfo{volume}{100}},
  \bibinfo{eid}{191901} (\bibinfo{year}{2012}).

\bibitem[{\citenamefont{Kadic et~al.}(2013)\citenamefont{Kadic, B\"{u}ckmann,
  Schittny, and Wegener}}]{Wegener_reviewRPP2008}
\bibinfo{author}{\bibfnamefont{M.}~\bibnamefont{Kadic}},
  \bibinfo{author}{\bibfnamefont{T.}~\bibnamefont{B\"{u}ckmann}},
  \bibinfo{author}{\bibfnamefont{R.}~\bibnamefont{Schittny}}, \bibnamefont{and}
  \bibinfo{author}{\bibfnamefont{M.}~\bibnamefont{Wegener}},
    \emph{\bibinfo{title}{Metamaterials beyond electromagnetism}},
  \bibinfo{journal}{Rep. Prog. Phys.}
  \textbf{\bibinfo{volume}{76}}, \bibinfo{pages}{126501}
  (\bibinfo{year}{2013}).




\bibitem[{\citenamefont{Oftadeh et~al.}(2014)\citenamefont{Oftadeh, Haghpanah,
  Vella, Boudaoud, and Vaziri}}]{Oftadeh_PRL2014}
\bibinfo{author}{\bibfnamefont{R.}~\bibnamefont{Oftadeh}},
  \bibinfo{author}{\bibfnamefont{B.}~\bibnamefont{Haghpanah}},
  \bibinfo{author}{\bibfnamefont{D.}~\bibnamefont{Vella}},
  \bibinfo{author}{\bibfnamefont{A.}~\bibnamefont{Boudaoud}}, \bibnamefont{and}
  \bibinfo{author}{\bibfnamefont{A.}~\bibnamefont{Vaziri}},
      \emph{\bibinfo{title}{Optimal Fractal-Like Hierarchical Honeycombs}},
  \bibinfo{journal}{Phys. Rev. Lett.} \textbf{\bibinfo{volume}{113}},
  \bibinfo{pages}{104301} (\bibinfo{year}{2014}).

 \bibitem[{\citenamefont{Wang et~al.}(2014)\citenamefont{Wang, Casadei, Shan,
  Weaver, and Bertoldi}}]{Wang_PRL2014}
\bibinfo{author}{\bibfnamefont{P.}~\bibnamefont{Wang}},
  \bibinfo{author}{\bibfnamefont{F.}~\bibnamefont{Casadei}},
  \bibinfo{author}{\bibfnamefont{S.}~\bibnamefont{Shan}},
  \bibinfo{author}{\bibfnamefont{J.~C.} \bibnamefont{Weaver}},
  \bibnamefont{and} \bibinfo{author}{\bibfnamefont{K.}~\bibnamefont{Bertoldi}},
  \emph{\bibinfo{title}{Harnessing Buckling to Design Tunable Locally Resonant Acoustic Metamaterials}},
  \bibinfo{journal}{Phys. Rev. Lett.} \textbf{\bibinfo{volume}{113}},
  \bibinfo{pages}{014301} (\bibinfo{year}{2014}).



 \bibitem[{\citenamefont{Overvelde et~al.}(2014)\citenamefont{Overvelde and Bertoldi}}]{Overvelde_JMPS2014}
\bibinfo{author}{\bibfnamefont{J.~T.~B.} \bibnamefont{Overvelde}} \bibnamefont{and}
  \bibinfo{author}{\bibfnamefont{K.}~\bibnamefont{Bertoldi}},
        \emph{\bibinfo{title}{Relating pore shape to the non-linear response of periodic elastomeric structures}},
  \bibinfo{journal}{J. Mech. Phys. Solids} \textbf{\bibinfo{volume}{64}},
  \bibinfo{pages}{351} (\bibinfo{year}{2014}).

\bibitem[{\citenamefont{Shim et~al.}(2012)\citenamefont{Shim, Perdigou, Chen,
  Bertoldi, and Reis}}]{Shim_PNAS2012}
\bibinfo{author}{\bibfnamefont{J.}~\bibnamefont{Shim}},
  \bibinfo{author}{\bibfnamefont{C.}~\bibnamefont{Perdigou}},
  \bibinfo{author}{\bibfnamefont{E.~R.} \bibnamefont{Chen}},
  \bibinfo{author}{\bibfnamefont{K.}~\bibnamefont{Bertoldi}}, \bibnamefont{and}
  \bibinfo{author}{\bibfnamefont{P.~M.} \bibnamefont{Reis}},
  \emph{\bibinfo{title}{Buckling-induced encapsulation of structured elastic shells under pressure}},
  \bibinfo{journal}{Proc. Natl. Ac. Sc. U.S.A.}
  \textbf{\bibinfo{volume}{109}}, \bibinfo{pages}{5978} (\bibinfo{year}{2012}).

\bibitem[{\citenamefont{Babaee et~al.}(2013)\citenamefont{Babaee, Shim, Weaver,
  Patel, and Bertoldi}}]{Babaee_AdvM2013}
\bibinfo{author}{\bibfnamefont{S.}~\bibnamefont{Babaee}},
  \bibinfo{author}{\bibfnamefont{J.}~\bibnamefont{Shim}},
  \bibinfo{author}{\bibfnamefont{J.~C.} \bibnamefont{Weaver}},
  \bibinfo{author}{\bibfnamefont{N.}~\bibnamefont{Patel}}, \bibnamefont{and}
  \bibinfo{author}{\bibfnamefont{K.}~\bibnamefont{Bertoldi}},
  \emph{\bibinfo{title}{3D Soft Metamaterials with Negative Poisson{\textquoteright}s Ratio}},
  \bibinfo{journal}{Adv. Mater.} \textbf{\bibinfo{volume}{25}},
  \bibinfo{pages}{5044} (\bibinfo{year}{2013}).



\bibitem{NoteEta} For the details about the calculation of $\eta$ in the simple case of uniaxial compression, see Supplemental Materials.







  \bibitem[{\citenamefont{Cho et~al.}(2014)\citenamefont{Cho, Shin, Costa, Kim,
  Kunin, Li, Lee, Yang, Han, Choi et~al.}}]{Cho_PNAS2014}
\bibinfo{author}{\bibfnamefont{Y.}~\bibnamefont{Cho}},
  \bibinfo{author}{\bibfnamefont{J.-H.} \bibnamefont{Shin}},
  \bibinfo{author}{\bibfnamefont{A.}~\bibnamefont{Costa}},
  \bibinfo{author}{\bibfnamefont{T.~A.} \bibnamefont{Kim}},
  \bibinfo{author}{\bibfnamefont{V.}~\bibnamefont{Kunin}},
  \bibinfo{author}{\bibfnamefont{J.}~\bibnamefont{Li}},
  \bibinfo{author}{\bibfnamefont{S.~Y.} \bibnamefont{Lee}},
  \bibinfo{author}{\bibfnamefont{S.}~\bibnamefont{Yang}},
  \bibinfo{author}{\bibfnamefont{H.~N.} \bibnamefont{Han}},
  \bibinfo{author}{\bibfnamefont{I.-S.} \bibnamefont{Choi}}, \bibnamefont{et~al.},
  \emph{\bibinfo{title}{Engineering the shape and structure of materials by fractal cut}},
  \bibinfo{journal}{Proc. Natl. Ac. Sc. U.S.A.}  (\bibinfo{year}{2014}).


\bibitem[{\citenamefont{Lakes}(1993)\citenamefont{Lakes}}]{Lakes_Nature1993}
\bibinfo{author}{\bibfnamefont{R.} \bibnamefont{Lakes}},
      \emph{\bibinfo{title}{Materials with Structural Hierarchy}},
  \bibinfo{journal}{Nature} \textbf{\bibinfo{volume}{361}},
  \bibinfo{pages}{511} (\bibinfo{year}{1993}).


\end{thebibliography}



\clearpage\subsection{Numerical Method}

We perform implicit finite element simulations with the commercial software Abaqus. For the 3D simulations (see fig. 2 of the main text), 
we use hybrid quadratic brick elements (Abaqus type C3D20H). For the 2D simulations, (see figs. 3, 4 and 5 of the main text), 
we use 2D quadratic plane strain elements (Abaqus type CPE8). For the plain beams, the elements are quadrilateral and for the metabeams, they are triangular. 
Metabeams with an even number of holes across their thickness are asymmetric. Therefore, in order to run robust simulations with fully symmetric initial geometries, 
we only consider metabeams with an odd number of holes across their width. To capture the instability without seeding the material with imperfections, we develop a two-steps protocol. First we perform a nonlinear stability analysis
to determine the bifurcation point ($u_b$, $F_b$) with a relative accuracy of $10^{-3}$.
%
%
Second, to probe the post buckling branch for $F>F_b$, we apply a transverse perturbation at the center of the beam and
release it once the beam is in the bifurcated ---buckled--- state. We then probe the buckled branch by smoothly
increasing the compression, $u$, to $3 u_b$. This protocol allows to
determine the location of the instability and the post-buckling behavior with high accuracy.

\subsection{Nonlinear Stiffening of Hyper-Elastic Beams.}

We consider a rubber beam under uniaxial loading in the direction $z$ and no deformations in the $y$ direction ---Fig.~\ref{fig:ax_uniax}. 
The strain energy density for incompressible neo-Hookean materials is~\cite{ogden}
\begin{equation}
 W=\frac{E}{6}\left(\lambda_x^2+\lambda_y^2+\lambda_z^2-3\right),
\end{equation}
where $\lambda_i$, are the stretches in the principal directions $i$ (length ratio before/after deformation), $E$ is the Young's modulus.
The incompressibility assumption translates into the equation $\lambda_x \lambda_y \lambda_z=1$.
For homogeneous deformations under constraints, the equilibrium state of the sample is simply given by
\begin{equation}
\sigma_i=\lambda_i\frac{\partial W}{\partial \lambda_i}-\Pi,\label{eq:equilibrium}
 \end{equation}
where $\sigma_i$ is the Cauchy stress and $\Pi$ is a Lagrange Multiplier~\cite{ogden}.
Since the deformations are spatially homogeneous, the $\lambda_i$ are constant in the sample and $\lambda_z$ relates then
simply to the vertical nominal strain $\tilde{\varepsilon}_{zz}$, by $\lambda_z=1+\tilde{\varepsilon}_{zz}$.

Here, we consider a uniaxial compression in the $z$ direction of a beam with no stresses in the $x$ directions and no deformations in the $y$ direction, therefore, 
the transverse stress $\sigma_x=0$ and the transverse stretch $\lambda_y=1$. We thus obtain
\begin{equation}
\Pi=\lambda_x\frac{\partial W}{\partial \lambda_x}\label{eq:Lagrange}
 \end{equation}
Therefore, the third equilibrium equation gives
 \begin{eqnarray}
\sigma_z&=&\lambda_z\frac{\partial W}{\partial \lambda_z}-\lambda_x\frac{\partial W}{\partial \lambda_x}\\
&=& \frac{E}{3}(\lambda_z^2-\lambda_x^{-2}).
 \end{eqnarray}
Using the incompressibility condition and the hypothesis $\lambda_y=1$, we obtain $\lambda_x=\lambda_z^{-1}$, so that
\begin{equation}
 \sigma_z=\frac{E}{3}\left(\lambda_z^2-\frac{1}{\lambda_z^2}\right).
\end{equation}
Finally, since the Cauchy stress and the stretch relate to the nominal 
stress as $\tilde{\sigma}_{zz}=\sigma_z/\lambda_z$~\cite{ogden}, we can write
\begin{equation}
 \tilde{\sigma}_{zz} = \frac{E}{3}\left(\lambda_z-\frac{1}{\lambda_z^2}\right)
 =\frac{E}{3}\left(1+\tilde{\varepsilon}_{zz}-\frac{1}{(1+\tilde{\varepsilon}_{zz})^3}\right)\label{eq:const1}.
\end{equation}
Therefore,
\begin{equation}
  \frac{\tilde{\sigma}_{zz}}{E}=\frac{1}{3}(1+\tilde{\varepsilon}_{zz}-(1+\tilde{\varepsilon}_{zz})^{-3}).
\end{equation}
This result is a good approximation for the pre-buckling stage of our 3D beams, although in 
the experiment, the hypothesis $\lambda_y=1$ is not strictly true.
\begin{figure}[t!]
    \includegraphics[width=0.8\columnwidth]{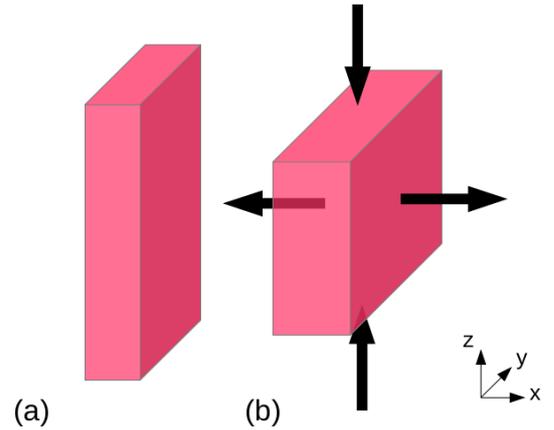}
  \caption{Beam under uniaxial compression in the undeformed (left) and deformed (right) states with no deformation in the $y$ direction (also called plane strain conditions). 
  The deformation are homogeneous and 
  boundary conditions impose $\sigma_x=0$ and $\lambda_y=0$.}
  \label{fig:ax_uniax}
  \end{figure}

%
%
\clearpage

In the following document we provide details for the 5 Movies accompanying the paper {\em Discontinuous Buckling of Wide Beams and Metabeams}.

\section{Experiment: Metabeam}

In Figs.~4ab, we show the force curves and snapshots for buckling experiments on several metabeams. The movie ({\em Experiment\_metabeam.mp4}) shows the buckling experiment with pictures 
(left) and the force curve (right) for a beam with $e=0.3$ and $\ell=0.3$. We clearly see that the microscopic structure 
changes upon buckling.

\section{Simulations: Metabeam}

The movies {\em Simulation\_metabeam1.avi} and {\em Simulation\_metabeam2.avi} show plane strain simulations of two metabeams 
of aspect ratio $5.4\%$, $e=0.1$ and $\ell=0.05$ and $0.70$ respectively. 
The movies {\em Simulation\_metabeam1\_zoom.avi} and {\em Simulation\_metabeam2\_zoom.avi} show the same simulation, but zoom on the middle of the beam.
For the beam with $\ell=0.7$, the microstructure hardly changes upon buckling. However, for $\ell=0.05$, the shape of the holes 
changes significantly in the transverse direction: in 
the compressed part (left), the elastic filaments are bent, while in the extended part (right), they are stretched.

\end{document}